# Prüfer transformation and its application to the numerical description of the motion of quantum particles with various spins in the fields of classical black holes

V. P.Neznamov[*], I.I.Safronov[†], V.E.Shemarulin[‡]

FSUE RFNC-VNIIEF, Russia, Sarov, Mira pr., 37, 607188

Abstract

Stable and reliable numerical integration of second-order radial equations in the fields of classical black holes can be performed by using the Prüfer transformation and by passing to the use of phase functions, which allows uniquely selecting the solutions with physical asymptotics in numerical calculations.

[*] vpneznamov@vniief.ru, vpneznamov@mail.ru  
[†] IISafronov@vniief.ru  
[‡] VEShemarulin@vniief.ru

# 1. Introduction

The self-adjoint second-order equation with spinor wave functions was used in [1] - [4] for a quantum mechanical description of the fermion motion in the fields of classical black holes. This equation was used in a flat space-time when solving quantum mechanics problems in attracting and repulsive Coulomb fields in [5], [6]. Quantum electrodynamics involving calculations of some physical effects based on a second-order equation with spinor wave functions was presented in [7], [8].

The absence of stationary states for scalar particles, photons, and fermions in the fields of classical black holes was shown in [9]. The reason is the singularity of the effective potential in a neighborhood of the event horizon, which leads to the regime of the particle "fall" on the event horizons, which is unacceptable for quantum mechanics [10] - [12]. The degenerate stationary states of scalar particles, photons, and fermions with energies $E_{st}$ are exceptions. The wave functions of these states vanish at the event horizon.

In this paper, we study the numerical solutions using the definition of the wave functions of degenerate stationary states for fermions, photons, and scalar particles in the fields of black holes. We prove the uniqueness of selection of the numerical solutions with the physical asymptotics by applying the Prüfer transformation and passing to the use of phase functions [13], [14].

# 2. Stationary solutions of second-order fermion equations in fields of classical black holes

We consider the Schwarzschild (S), Reissner-Nordström (RN), Kerr (K), and Kerr-Newman (KN) black holes.

The stationary Kerr-Newman metric is characterized by a point-like source with the mass $M$ and charge $Q$ rotating with the angular momentum $J = Mca$. We have $Q = 0$ for the Kerr metric, $J = 0$ for the Reissner-Nordström metric, and $Q = 0, J = 0$ for the Schwarzschild metric.

We use the following notation: $r_0 = 2GM/c^2$ is the gravitational radius of the Schwarzschild field, $r_\pm$ are the radii of the outer and inner event horizons (for the Schwarzschild metric $r_+ = r_0, r_- = 0$), $G$ is the gravitational constant, $c$ is the speed of light, and $r_Q = \sqrt{G}Q/c^2$ for the Reissner-Nordström and Kerr-Newman metrics.

We also use the dimensionless variables $\varepsilon = E/mc^2$ and $\rho = r/l_c$, where $E$ and $m$ are the energy and mass of a quantum particle, $l_c = \hbar/mc$ is its Compton wavelength, and

$$\alpha = \frac{r_0}{2l_c} = \frac{GMm}{\hbar c} = \frac{Mm}{M_P^2}, \quad \alpha_Q = \frac{r_Q}{l_c} = \frac{\sqrt{G}Qm}{\hbar c} = \frac{\sqrt{\alpha_{fs}}}{M_P}m\frac{Q}{e}, \quad \alpha_a = \frac{a}{l_c}, \quad \alpha_{em} = \frac{qQ}{\hbar c} = \alpha_{fs}\frac{qQ}{e^2}, \text{ where}$$

$M_P = \sqrt{\frac{\hbar c}{G}} = 2.2 \cdot 10^{-5}$ g $\left(1.2 \cdot 10^{19} \text{ GeV}/c^2\right)$ is the Planck mass, $\alpha_{fs} = \frac{e^2}{\hbar c} \approx \frac{1}{137}$ is the electromagnetic fine structure constant, $\alpha, \alpha_{em}$ are the gravitational and electromagnetic coupling constants, and $\alpha_Q$ and $\alpha_a$ are the dimensionless constants characterizing the source of the electromagnetic field with the charge $Q$ in the Reissner-Nordström and Kerr-Newman metrics and the ratio of the angular momentum $J$ to $Mc$ in the Kerr and Kerr-Newman metrics.

Below, we use the system of units with $\hbar = c = 1$.

After separating the variables, the second-order equation for the radial spinor function can be written in the form [1] - [8]:

$$\frac{d^2\psi}{d\rho^2} + 2\left(E_{Schr} - U_{eff}\right)\psi = 0, \tag{1}$$

where $E_{Schr} = \left(\varepsilon^2 - 1\right)/2$. The explicit form of the effective potentials $U_{eff}$ for the considered metrics was derived and presented in [2] - [4], [9].

The function $\omega(\rho, \varepsilon)$ depending on the particle energy appears in the effective potentials[§]

$$\omega_{KN} = \varepsilon\left(1 + \frac{\alpha_a^2}{\rho^2}\right) - \frac{\alpha_a m_\varphi}{\rho^2} - \frac{\alpha_{em}}{\rho}, \tag{2}$$

$$\omega_K = \varepsilon\left(1 + \frac{\alpha_a^2}{\rho^2}\right) - \frac{\alpha_a m_\varphi}{\rho^2}, \tag{3}$$

$$\omega_{RN} = \varepsilon - \frac{\alpha_{em}}{\rho}, \tag{4}$$

$$\omega_S = \varepsilon, \tag{5}$$

where $m_\varphi = -j, -j+1, ..., j$ is the projection of the total momentum $j$.

If $\omega(\rho_\pm, \varepsilon) \neq 0$ on the event horizon, then the regime of the particle "fall" on the event horizons, which is unacceptable for quantum mechanics, exists for all black holes [10] - [12]. Here, $(\rho_\pm)_{KN} = \alpha \pm \sqrt{\alpha^2 - \alpha_Q^2 - \alpha_a^2}$, $(\rho_\pm)_K = \alpha \pm \sqrt{\alpha^2 - \alpha_a^2}$, $(\rho_\pm)_{RN} = \alpha \pm \sqrt{\alpha^2 - \alpha_Q^2}$, $(\rho_+)_S = 2\alpha$, $(\rho_-)_S = 0$.

---

[§] It is a parameter for the Schwarzschild metric.

In the case $\omega(\rho_\pm, \varepsilon) = 0$, there is a potential well in the neighborhood of the event horizons for all black holes:

$$U_{eff}\big|_{\rho \to \rho_\pm} = -\frac{3}{32}\frac{1}{(\rho - \rho_\pm)^2}. \tag{6}$$

Here, the coefficient $\frac{3}{32} < \frac{1}{8}$, which indicates the absence of the regime of particle "fall" on the event horizons.

According to (2) - (5), the energies of the stationary states for the different black holes are

$$\varepsilon_{KN} = \frac{\alpha_a m_\varphi + \alpha_{em}}{\alpha_a^2 + (\rho_\pm)^2}, \tag{7}$$

$$\varepsilon_K = \frac{\alpha_a m_\varphi}{\alpha_a^2 + (\rho_\pm)^2}, \tag{8}$$

$$\varepsilon_{RN} = \frac{\alpha_{em}}{\rho_\pm}, \tag{9}$$

$$\varepsilon_S = 0. \tag{10}$$

The asymptotics of the solutions of Eq. (1) as $\rho \to \infty$ are given by

$$\psi\big|_{\rho \to \infty} = C_1 \chi_1(\rho) e^{-\sqrt{1-\varepsilon^2}\rho} + C_2 \chi_2(\rho) e^{\sqrt{1-\varepsilon^2}\rho}, \tag{11}$$

where $\chi_1(\rho), \chi_2(\rho)$ are power-low functions of $\rho$.

The asymptotics of solutions of Eq. (1) as $\rho \to \rho_\pm$, in view of the leading singularity of the effective potential (see (6)), are given by

$$\psi\big|_{\rho \to \rho_\pm} = C_3 |\rho - \rho_\pm|^{1/4} + C_4 |\rho - \rho_\pm|^{3/4}. \tag{12}$$

For stationary bound states with exponentially decreasing solutions as $\rho \to \infty$ and with the asymototics of the solutions $\psi \sim |\rho - \rho_\pm|^{1/4}$ in a neighborhood of the event horizons (see [2] - [4]), we have to set the constants $C_2$ and $C_4$ equal to zero in (11), (12), which gives rise to two questions. Are there solutions of Eq. (1) of this form? Is it possible to uniquely select such solutions in numerical calculations?

*2.1 A model second-order equations*

To answer the above questions, we consider the model equation reflecting the basic features of Eq. (1) as $\rho \to \infty$ and $\rho \to \rho_\pm$:

$$\frac{d^2\psi}{d\rho^2} + \frac{3}{16\rho^2}\psi + (\varepsilon^2 - 1)\psi = 0. \tag{13}$$

Equation (13) was proposed for the analysis by M. N. Smolyakov in our joint discussion. Asymptotics (11), (12) appear in Eq. (13) if $|\rho - \rho_{\pm}|$ is replaced with $\rho$.

In the case $\varepsilon = 0$ (the Schwarzschild metric, see (10)), Eq. (13) has the analytical solution

$$\psi(\rho) = C_3 \sqrt{\rho} K_{1/4}(\rho) + C_4 \sqrt{\rho} I_{1/4}(\rho). \tag{14}$$

Here, $I_{1/4}(\rho)$ and $K_{1/4}(\rho)$ are the modified Bessel function and Macdonald functions with the asymptotics

$$\sqrt{\rho} K_{1/4}(\rho) : \begin{cases} e^{-\rho}, & \rho \to \infty, \\ \rho^{1/4}, & \rho \to 0, \end{cases} \tag{15}$$

$$\sqrt{\rho} I_{1/4}(\rho) : \begin{cases} e^{\rho}, & \rho \to \infty, \\ \rho^{3/4}, & \rho \to 0. \end{cases} \tag{16}$$

We see that if $C_4 = 0$ in Eq. (14), then there remains the sought solution for the stationary bound state $\varepsilon = 0$, which exponentially decreases as $\rho \to \infty$ and is $\sim \rho^{1/4}$ as $\rho \to 0$.

In the case $\varepsilon \neq 0$ (see Eqs. (7) - (9)), solutions of Eq. (13) have asymptotics (11) as $\rho \to \infty$ and asymptotics (15), (16) as $\rho \to 0$. Obviously, for stationary bound states with energies (7) - (9), there are also eigenfunctions with the asymptotics $\sim \rho^{1/4}$ as $\rho \to 0$ and an exponential decrease as $\rho \to \infty$. Below, we present such solutions obtained in numerical calculations. With reasonable confidence, we can make a similar conclusion for the solutions of Eq. (1). For the final proof of this conclusion and for solving the problem of selecting the solutions with physical asymptotics in the numerical calculations, we resort to phase functions in quantum theory [13], [14], [2] - [4].

### 2.2 The Pruefer transformation and phase functions

We apply the Prüfer transformation to Eq. (1).

Let

$$\psi(\rho) = P(\rho) \sin \Phi(\rho), \tag{17}$$

$$\frac{d\psi(\rho)}{d\rho} = P(\rho) \cos \Phi(\rho). \tag{18}$$

Then

$$\psi \bigg/ \frac{d\psi(\rho)}{d\rho} = \tan \Phi(\rho) \tag{19}$$

and Eq. (1) can be written as

$$\frac{d\Phi}{d\rho} = \cos^2 \Phi + 2(E_{Schr} - U_{eff}) \sin^2 \Phi, \tag{20}$$

$$\frac{d \ln P}{d\rho} = \left(1 - 2\left(E_{Schr} - U_{eff}\right)\right)\sin\Phi\cos\Phi. \tag{21}$$

Equation (21) must be solved after finding the eigenvalues $\varepsilon_n$ and eigenfunctions $\Phi_n(\rho)$ from Eq. (20). The probability density of detecting particles at a distance $\rho$ in a spherical layer $d\rho$ is

$$w(\rho) = P_n^2(\rho)\sin^2\Phi_n(\rho). \tag{22}$$

*2.2.1 Asymptotics of the functions $\Phi(\rho), P(\rho),$ and $\psi(\rho)$.*

The asymptotics of the solutions to Eqs. (20), (21) and asymptotics (11), (12) for the original equation (1) are given in Table 1.

Table 1. Asymptotics of Eqs. (1), (20), (21).

| № | | $\Phi(\rho)$ | $(-1)^k \sin\Phi(\rho)$ | $P(\rho)$ | $\psi(\rho)$ |
|---|---|---|---|---|---|
| 1 | $\rho \to \infty$ | $-\arctan\dfrac{1}{\sqrt{1-\varepsilon^2}} + k\pi$ | $\dfrac{-1}{\sqrt{2-\varepsilon^2}}$ | $C_1'\chi_1(\rho)e^{-\sqrt{1-\varepsilon^2}\rho}$ | $C_1\chi_1(\rho)e^{-\sqrt{1-\varepsilon^2}\rho}$ |
| 2 | $\rho \to \infty$ | $+\arctan\dfrac{1}{\sqrt{1-\varepsilon^2}} + k\pi$ | $\dfrac{1}{\sqrt{2-\varepsilon^2}}$ | $C_2'\chi_2(\rho)e^{\sqrt{1-\varepsilon^2}\rho}$ | $C_2\chi_2(\rho)e^{\sqrt{1-\varepsilon^2}\rho}$ |
| 3 | $\rho \to \rho_\pm$ | $4\lvert\rho-\rho_\pm\rvert + k\pi$ | $4\lvert\rho-\rho_\pm\rvert$ | $C_3'\lvert\rho-\rho_\pm\rvert^{-3/4}$ | $C_3\lvert\rho-\rho_\pm\rvert^{1/4}$ |
| 4 | $\rho \to \rho_\pm$ | $\dfrac{4}{3}\lvert\rho-\rho_\pm\rvert + k\pi$ | $\dfrac{4}{3}\lvert\rho-\rho_\pm\rvert$ | $C_4'\lvert\rho-\rho_\pm\rvert^{-1/4}$ | $C_4\lvert\rho-\rho_\pm\rvert^{3/4}$ |
| 5 | $\rho \to 0$ | $\dfrac{2}{3}\rho + k\pi$ | $\dfrac{2}{3}\rho$ | $C_5'\rho^{1/2}$ | $C_5\rho^{3/2}$ |
| 6 | $\rho \to 0$ | $-2\rho + k\pi$ | $-2\rho$ | $C_6'\rho^{-3/2}$ | $C_6\rho^{-1/2}$ |

The first, third, and fifth rows of this Table correspond to the asymptotics of the solutions of Eq. (1) for stationary bound states with energies (7) - (10). The aymptotics in the first and third rows correspond to the domain of wave functions $\rho \in [\rho_+, \infty)$; the asymptotics in the third and fifth rows correspond to the region $\rho \in (0, \rho_-]$. The analysis as $\rho \to 0$ is applicable only for the Reissner-Nordström metric. In that case, $\left.(U_{eff})_{RN}\right|_{\rho \to 0} = 3/8\rho^2$ [3]. The effective potential for the Kerr and Kerr-Newman metrics is finite at $\rho = 0$, and we select the solutions with the physical asymptotics using the two-sheet topology of the considered metrics (see paper [4]).

*2.2.2 Singular points of the equations for phase function*

1. The point $\rho = 0$.

The analysis is applicable only to the Reissner-Nordström metric (see Sec. 2.2.1). There is the singularity $U_{Eff}\big|_{\rho\to 0} = \dfrac{3}{8}\dfrac{1}{\rho^2}$ in a neighborhood of the point $\rho = 0$. Therefore, Eq. (20) for the phase function can be written in the form

$$\frac{d\Phi}{d\rho} = \frac{\rho^2 \cos^2\Phi + 2\left(\rho^2 E_{Schr} - \dfrac{3}{8} + O(\rho)\right)\sin^2\Phi}{\rho^2}. \qquad (23)$$

The singular points can be found from the conditions that the numerator and denominator of the right-hand side of the last equation vanish. Consequently, the following points on the ordinate axis are singular:

$$(0, \Phi_k), \quad \Phi_k = k\pi, \quad k = 0, \pm 1, \pm 2, ...$$

Substituting the relation $\Phi(\rho) \cong A\rho$ in Eq. (20) we can easily find the linear asymptotics of solutions near the point $(\rho, \Phi) = (0, 0)$:

$$\left.\frac{d\Phi}{d\rho}\right|_{\rho=0,\,\Phi=0} = A = 1 - \frac{3}{4}A^2 \text{ и } A_{1,2} = -2, \frac{2}{3}.$$

2. Points $\rho = \rho_\pm$.

In a neighborhood of the points $\rho = \rho_\pm$, there is the singularity $U_{Eff}\big|_{\rho\to\rho_\pm} = -\dfrac{3}{32}\dfrac{1}{(\rho-\rho_\pm)^2}$. Finding other points in these cases is essentially similar to the preceding case $\rho = 0$. It follows that the following points are singular in this cases:

$$(\rho_\pm, \Phi_k), \quad \Phi_k = k\pi, \ k = 0, \pm 1, \pm 2, ...$$

The linear asymptotics of $\Phi(\rho) \cong A|\rho - \rho_\pm|$ of the solutions in a neighborhood of the points $(\rho, \Phi) = (\rho_\pm, 0)$ can be determined from the equation $\left.\dfrac{d\Phi}{d\rho}\right|_{\rho=\rho_\pm,\,\Phi=0} = A = 1 + \dfrac{3}{16}A^2$. Consequently, the angular coefficients are $A_{1,2} = 4, 4/3$.

3. The point $\rho = \infty$. The behavior of the effective potential $U_{eff}$ at infinity can be described by the asymptotic formula

$$U_{eff}\big|_{\rho\to\infty} = \frac{c}{\rho} + O\left(\frac{1}{\rho^2}\right), \quad c = \text{const.} \qquad (24)$$

The transformation of $\xi = 1/\rho$ takes the point $\rho = \infty$ to zero and transforms Eq. (20) into the equation

$$-\xi^2 \frac{d\Phi}{d\xi} = \cos^2 \Phi + 2(E_{Schr} - U_{eff})\sin^2 \Phi. \quad (25)$$

The point $\xi = 0$ on the phase plane $(\xi, \Phi)$ corresponds to singular points of the appropriate dynamical system that are located on the ordinate axis $(\xi, \Phi) = (0, \Phi_k^\pm)$, $k = 0, \pm 1, \pm 2, \ldots$ It follows from Eq. (25) that the asymptotic values of the phase $\Phi(\xi)|_{\xi \to 0}$ are $\Phi_k^\pm = \pm \text{arctg}\left(\frac{1}{\sqrt{1-\varepsilon^2}}\right) + k\pi$ at $\varepsilon \neq \pm 1$.

*2.2.3 The character of singular points of Eq. (20).*

The character of singular points can be determined according to the well-known algorithm (see [15], Chap. 6; [16], p. 441 - 442).

We present Eq. (23) in the form

$$\frac{d\Phi}{d\rho} = \frac{c\rho + d(\Phi - \Phi_k) + P_1(\rho, (\Phi - \Phi_k))}{a\rho + b(\Phi - \Phi_k) + Q_1(\rho, (\Phi - \Phi_k))},$$

$a = b = c = d = 0$, $P_1(\rho, (\Phi - \Phi_k))$, and $Q_1(\rho, (\Phi - \Phi_k))$ are second-order infinitesimals with respect to $\rho$ and $(\Phi - \Phi_k)$. The determinant vanishes,

$$\Delta_k = \begin{vmatrix} a & b \\ c & d \end{vmatrix} = 0,$$

and therefore, all points $(0, \Phi_k)$ are higher-order singular points.

Similar arguments show that the points $(\rho_\pm, \Phi_k)$, $(\infty, \Phi_k^-)$ and $(\infty, \Phi_k^+)$ are also higher-order singular points.

At the same time, the behavior of the integral curves of Eq. (20) in the neighborhoods of the singular points shows that the points $(0, \Phi_k)$ and $(\rho_\pm, \Phi_k)$ are singular points that combine the "node" and "saddle" singularity types, while the singular points $(\infty, \Phi_k^-)$ are the regular "saddle" ones, and the singular points $(\infty, \Phi_k^+)$ are the regular "node" points.

*2.2.4 Behavior of the integral curves $\Phi = \Phi(\rho, C)$ of Eq. (20) in a neighborhood of singular points of the effective potential.*

Equation (20) for the phase function can be written as a Riccati equation for $\tan\Phi$:

$$\frac{d\tan\Phi}{d\rho} = 1 + 2(E_{Schr} - U_{eff})\tan^2\Phi. \quad (26)$$

1. The singularity in a neighborhood of the point $\rho = 0$ is $U_{Eff}|_{\rho \to 0} = \frac{3}{8}\frac{1}{\rho^2}$, and Eq. (26)

can be written as

$$\frac{d\tan\Phi}{d\rho} = 1 - \frac{3}{4}\frac{1}{\rho^2}\tan^2\Phi. \tag{27}$$

The general solution of Eq. (27) is given by

$$\tan\Phi = \frac{2\rho(C\rho^2 + 1)}{3C\rho^2 - 1}, \quad C = \text{const.} \tag{28}$$

The family $\Phi = \Phi(\rho, C)$ of integral curves (28) is presented in Fig. 1. The upper separatrix $\Phi \approx 2/3\rho$ corresponds to the value $C = \pm\infty$ and the lower separatrix $\Phi \approx -2\rho$ corresponds to $C = 0$. The behavior of the integral curves can be characterized as the "node" type in a neighborhood of the lower separatrix and the "saddle" type above the upper separatrix. We can see that the integral curves touch the lower separatrix in a neighborhood of the point $\rho = 0$ in the "node" region.

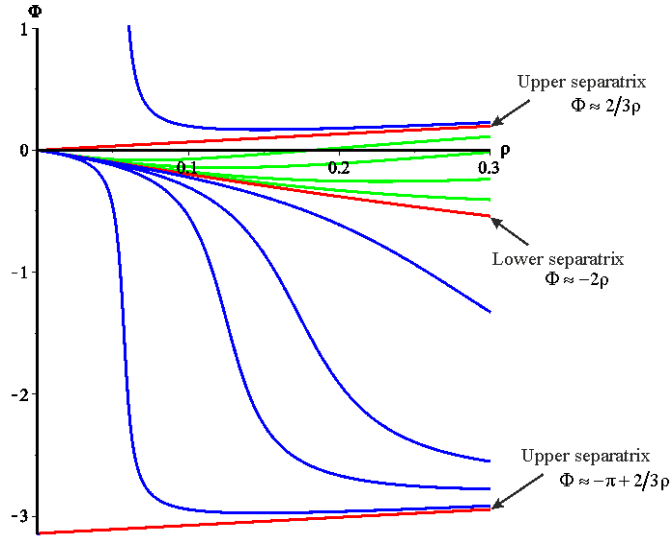

Fig. 1. Integral curves $\Phi = \Phi(\rho, C)$ in a neighborhood of the point $\rho = 0$ $-1 < \varepsilon < 1$.

In the case where two event horizons $\rho_\pm$ exist, we consider the solutions with $\omega(\rho_\pm) = 0$ (see Eqs. (2) - (5)).

2. The singularity in a neighborhood of the point $\rho = \rho_-$ is $U_{Eff}\big|_{\rho \to \rho_-} = -\frac{3}{32}\frac{1}{(\rho - \rho_-)^2}$

and Eq. (26) can be written as

$$\frac{d\tan\Phi}{d\rho} = 1 + \frac{3}{16}\frac{1}{(\rho - \rho_-)^2}\tan^2\Phi. \tag{29}$$

The general solution of Eq. (29) in the region $\rho < \rho_-$ has the form

$$\tan\Phi = \frac{4(\rho-\rho_-)\left(1-C\sqrt{\rho_--\rho}\right)}{1-3C\sqrt{\rho_--\rho}}, \quad C = \text{const.} \tag{30}$$

The family of integral curves $\Phi = \Phi(\rho-\rho_-, C)$ for the Reissner-Nordström metric is presented in Fig. 2. The the upper separtrix $\Phi \approx -4/3(\rho_- - \rho)$ corresponds to the value $C = \pm\infty$, the lower separatrix $\Phi \approx -4(\rho_- - \rho)$ corresponds to $C = 0$. As in the preceding case, the behavior of the integral curves in a neighborhood of the lower separatrix is of the "node" type, and the behavior above the upper separatrix is of the "saddle" type.

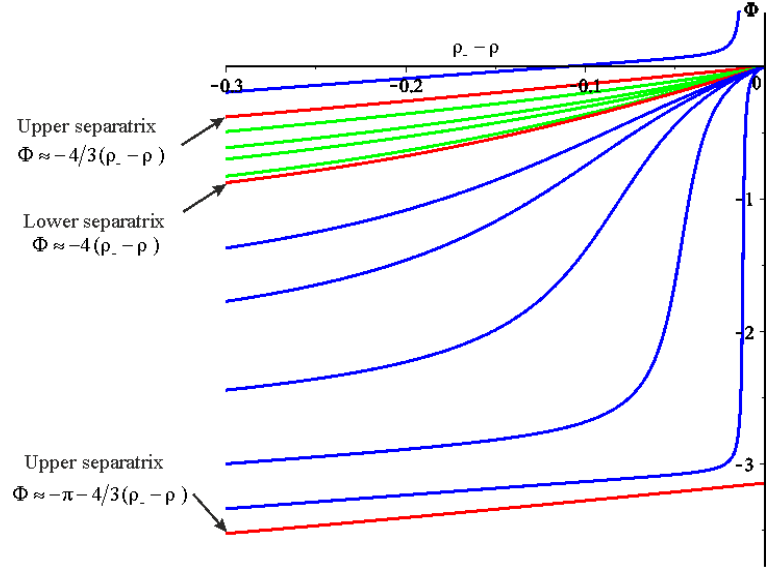

Fig. 2. Integral curves $\Phi = \Phi(\rho-\rho_-, C)$ in a neighborhood of the point $\rho = \rho_-$.

The detailed consideration of the behavior of the integral curves in a neighborhood of the point $\rho = \rho_-$ in the "node" region shows that all of them touch the lower separatrix (Fig. 3).

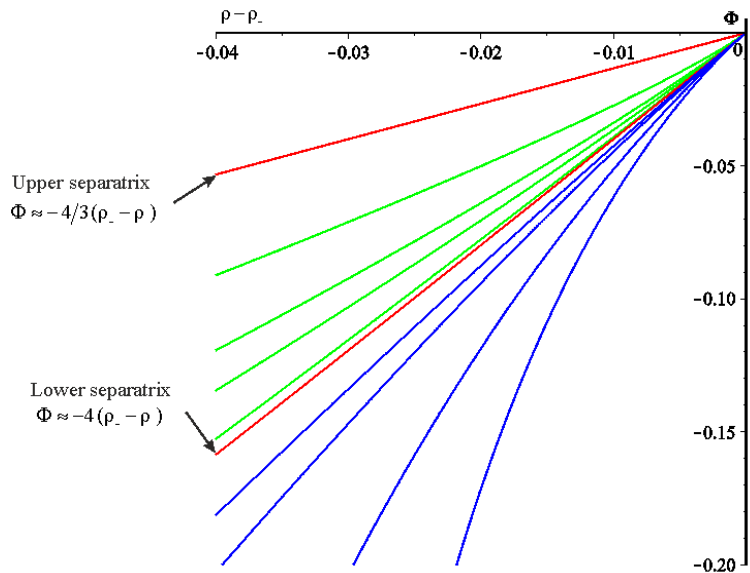

Fig. 3. Integral curves $\Phi = \Phi(\rho-\rho_-, C)$ in a small neighborhood of $\rho = \rho_-$.

3. The singularity in a neighborhood of the point $\rho = \rho_+$ is $U_{eff}\big|_{\rho \to \rho_+} = -\dfrac{3}{32}\dfrac{1}{(\rho-\rho_+)^2}$,

and Eq. (26) can be written as

$$\frac{d\tan\Phi}{d\rho} = 1 + \frac{3}{16}\frac{1}{(\rho-\rho_+)^2}\tan^2\Phi. \tag{31}$$

The general solution of Eq. (31) in the region $\rho > \rho_+$ has the form

$$\tan\Phi = \frac{4(\rho-\rho_+)\left(1-C\sqrt{\rho-\rho_+}\right)}{1-3C\sqrt{\rho-\rho_+}}, \quad C = \text{const.} \tag{32}$$

The family $\Phi = \Phi(\rho-\rho_+,C)$ of integral curves (32) for the Reissner-Nordström metric is presented in Fig. 4. The upper separatrix $\Phi \approx 4(\rho-\rho_+)$ corresponds to the value $C=0$, and the lower separatrix $\Phi \approx 4/3(\rho-\rho_+)$ corresponds to $C=\pm\infty$. In contrast to the two previous cases, the behavior of the integral curves in a neighborhood of the upper separatrix is of the "node" type, and the behavior below the lower separatrix is of the "saddle" type.

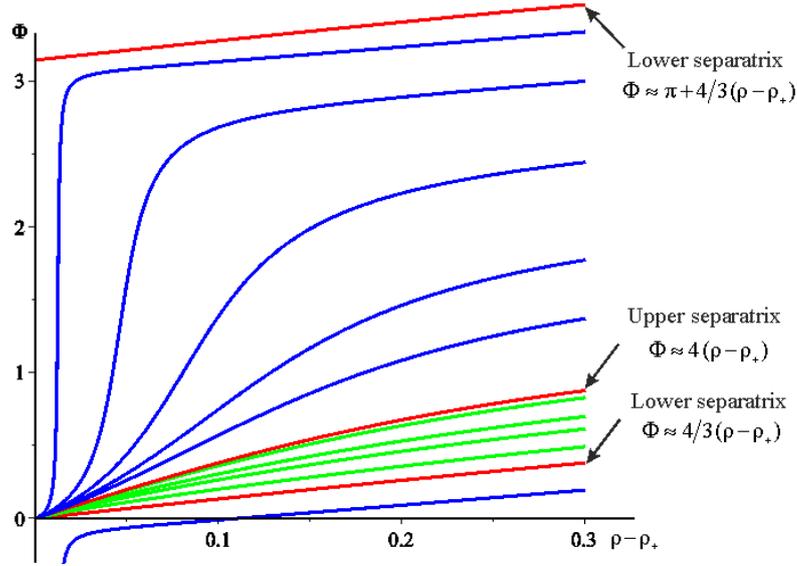

Fig. 4. Integral curves $\Phi = \Phi(\rho-\rho_+,C)$ in a neighborhood of $\rho = \rho_+$.

The detailed consideration of the behavior of the integral curves in a neighborhood of the point $\rho = \rho_+$ in the "node" region shows that all of them touch the upper separatrix (Fig. 5).

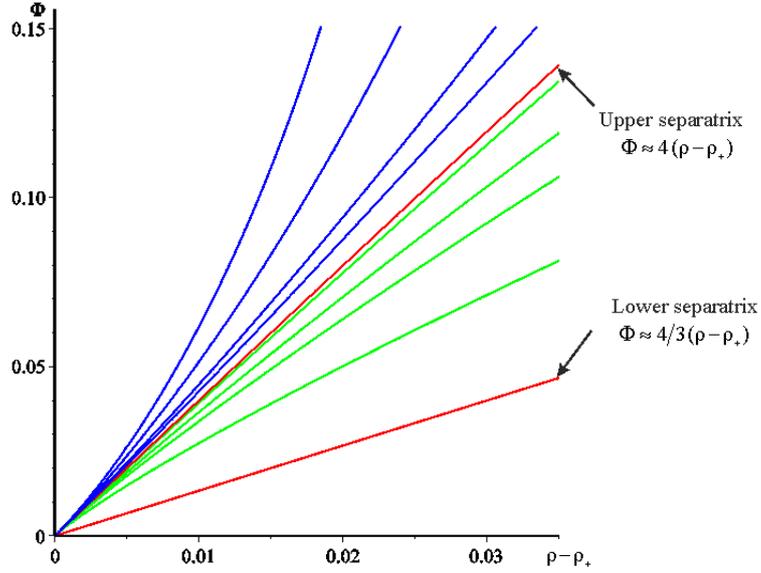

Fig. 5. Integral curves $\Phi = \Phi(\rho - \rho_+, C)$ in a small neighborhood of $\rho = \rho_+$.

4. The effective potential has the asymptotics of (24) in a neighborhood of the point $\rho = \infty$ for the considered fields, and Eq. (26) can be written as

$$\frac{d\tan\Phi}{d\rho} = 1 + c_1 \tan^2\Phi, \quad c_1 = 2E_{Schr} = \varepsilon^2 - 1 < 0. \tag{33}$$

The general solution of Eq. (33) is given by

$$\frac{1}{2\sqrt{-c_1}} \ln \frac{1 + \sqrt{-c_1}\tan\Phi}{1 - \sqrt{-c_1}\tan\Phi} = \rho + C, \quad \sqrt{-c_1}|\tan\Phi| < 1,$$

$$\frac{1}{2\sqrt{-c_1}} \ln \frac{\sqrt{-c_1}\tan\Phi + 1}{\sqrt{-c_1}\tan\Phi - 1} = \rho + C, \quad \sqrt{-c_1}|\tan\Phi| > 1, \ C = \text{const}. \tag{34}$$

It follows from (34) that for all solutions of Eq. (20) we have $\lim_{\rho \to \infty} \Phi = \arctan \frac{1}{\sqrt{1-\varepsilon^2}} + k\pi \mp 0$ (the minus sign refers to the first row in formula (34), and the plus sign to the second row).

The transformation $\xi = 1/\rho$ takes the solution $\Phi = \Phi(\rho, C)$ in Eq. (34) into the solution

$$\Phi = \Phi(\xi, C). \tag{35}$$

The family of integral curves (35) is presented in Fig. 6. The constant is $C = 0$ on the marked integral curves and $C > 0$ between them. The ordinate axis corresponds to the value $C = -\infty$.

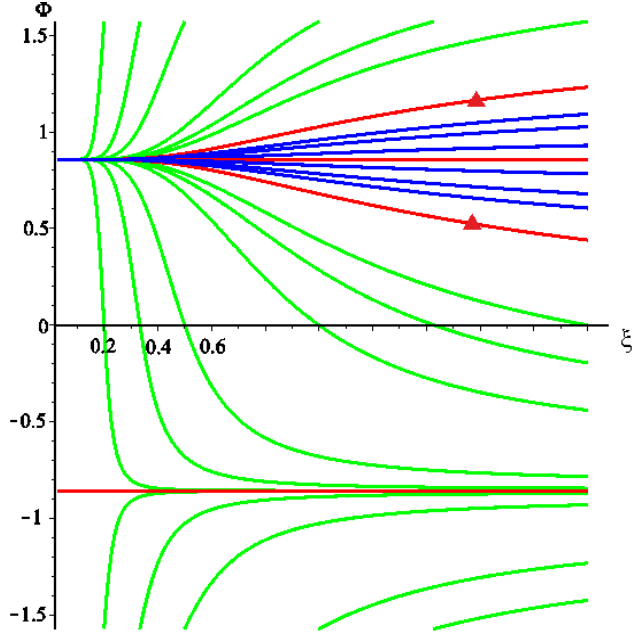

Fig. 6. Integral curves $\Phi = \Phi(\xi, C)$ in a neighborhood of the point at $\xi = 0$ ($\rho = \infty$) at $\varepsilon = 0.5$, $\Phi_k^\pm = \pm 0.8570719477$.

The detailed analysis of the behavior of the integral curves in a neighborhood of the point $\xi = 0$ in the "node" region (above the abscissa axis $\Phi = 0$) shows that all these curves asymptotically tend to the straight line $\Phi_0^+ = \arctan \dfrac{1}{\sqrt{1-\varepsilon^2}}$ corresponding to the value $C = +\infty$ and enter the "node" $\left(\xi = 0, \Phi_0^+\right)$ (see Fig. 6). As $\xi \to 0$, all curves from the neighborhood contained in the "saddle" region (below the abscissa axis) which, by definition, do not coincide with the "saddle" separatrix $\Phi_0^- = -\operatorname{arctg} \dfrac{1}{\sqrt{1-\varepsilon^2}}$ $(C = -\infty)$, move away from it and enter the "nodal region", and hence terminate at the "node" $\left(\xi = 0, \Phi_0^+\right)$.

Such behavior of the integral curves in a neighborhood of the point $\rho = \infty$ ($\xi = 0$) proves that it is a regular singular point.

*2.2.5 The main characteristics of singular points.*

1. The regular singular points $\left(\infty, \Phi_k^-\right)$ are "saddles" with the separatrices

$$\Phi_k^- = -\operatorname{arctg} \dfrac{1}{\sqrt{1-\varepsilon^2}} + k\pi, \ k = 0, \pm 1, \pm 2, \ldots$$

2. The regular singular points $\left(\infty, \Phi_k^+\right)$ are "nodes" with the separatrices

$$\Phi_k^+ = \operatorname{arctg} \dfrac{1}{\sqrt{1-\varepsilon^2}} + k\pi.$$

3. The singular points $\left(\rho_{\pm}, \Phi_k^{(1,2)}\right), \left(0, \Phi_k^{(3,4)}\right)$ are higher-order singular points combining singularities of the "node" and "saddle" types.

a) $\Phi_k^{(1)} = 4|\rho - \rho_{\pm}| + k\pi$ is a separatrix entering the "node" part of the singular point.

b) $\Phi_k^{(2)} = \frac{4}{3}|\rho - \rho_{\pm}| + k\pi$ is a separatrix separating the "saddle" part of the singular point from the "node" part. The "node" curves $\Phi_k^{(1)}$ are separated from the separatrix $\Phi_k^{(2)}$ at any $\rho \neq \rho_{\pm}$.

c) $\Phi_k^{(3)} = \frac{2}{3}\rho + k\pi$ is a separatrix separating the "saddle" part of the singular point from the "node" part.

d) $\Phi_k^{(4)} = -2\rho + k\pi$ is a separatrix entering the "node" part of the singular point. Here, the "node" curves $\Phi_k^{(4)}$ are also separated from the separatrix $\Phi_k^{(3)}$ at any $\rho \neq 0$.

*2.2.6 Strategy of integration the equations for phase function*

We must remember that the integration in numerical calculations must start in a neighborhood of a "saddle" separatrix and terminate in a neighborhood of a "node" separatrix. The backward integration is impossible because the exit from the "node" is unstable and entering the "saddle" with a single separatrix is impossible because of the finite accuracy of numerical calculations.

We consider the domain $\rho \in [\rho_+, \infty)$. In this case, for stationary bound states with the wave functions exponentially decreasing as $\rho \to \infty$, we have to integrate Eq. (20) "from right to left" (from $\infty$ to $\rho_+$) choosing $\Phi_k^-$ at the starting integration point and terminating the integration at the point $\Phi_k^{(1)}$. The integration from $\Phi_k^{(2)}$ to $\Phi_k^+$ leads to solutions that grow exponentially as $\rho \to \infty$.

For the domain $\rho \in (0, \rho_-]$, the solutions for stationary bound states can be obtained by integrating from $\Phi_k^{(3)}$ to the radius of the inner event horizon, i.e., to $\Phi_k^{(2)}$.

*2.2.7 Numerical solutions of the equations for phase functions*

A specialized code has been worked out for solving first-order nonlinear equations (20), (21) numerically. The Cauchy problem with a given initial condition is solved numerically for the allowed set of values of $\varepsilon$. The implicit fifth-order Runge-Kutta method with the step control is used in solving this problem (the Ehle scheme of three-stage Radau II A metod [17]).

The "right-to-left" or "left-to-right" integration directions can be determined in accordance with Sec. 2.2.6.

Numerous results of calculations using this code are presented in [2] - [6]. The electron energy levels of hydrogen-like atoms and probability densities of electron energy states are numerically simulated in [5].

The comparison of the normalized probability densities of analytical and numerical solutions of the model equation (13) with $\varepsilon = 0$ is demonstrated in Fig. 7. We see that they completely coincide with each other.

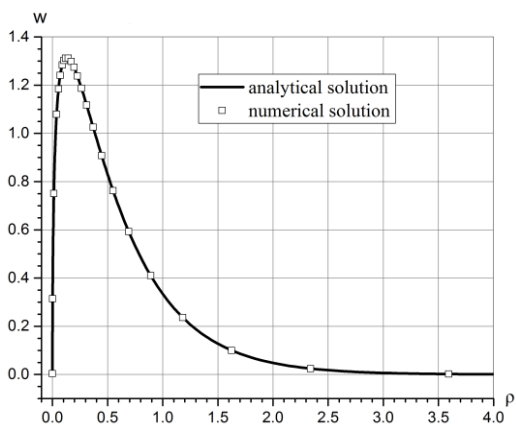

Fig. 7. The dependences $w(\rho)$ at $\varepsilon = 0$

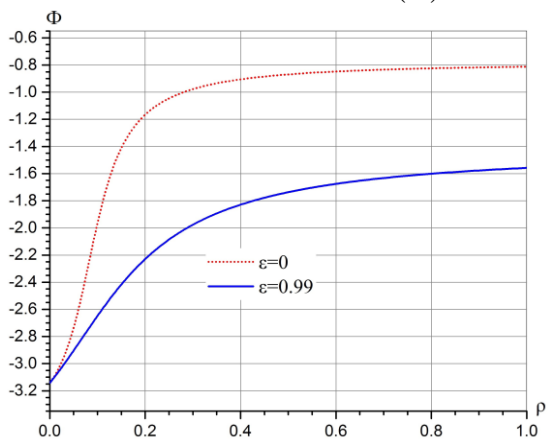

Fig. 8. The dependences $\Phi(\rho, \varepsilon)$.

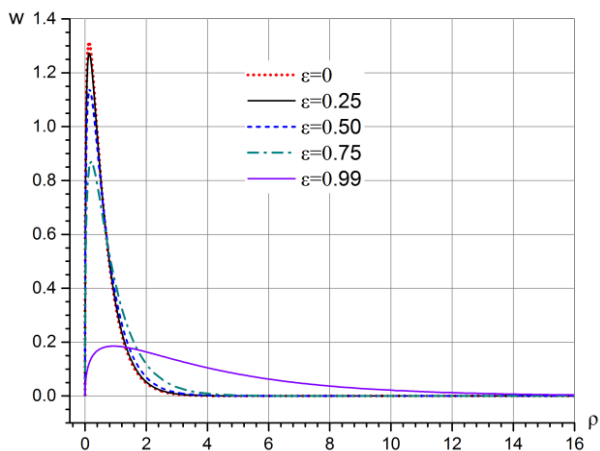

Fig. 9. The dependences $w(\rho)$ at $0 \leq \varepsilon < 1$.

The comparison of the numerical solutions of Eqs. (13) with different values of $0 \leq \varepsilon < 1$ is demonstrated in Figs. 8, 9. The curves $\Phi(\rho)$ obviously defer from each other as $\rho \to \infty$ in accordance with the initial condition (see row 1 of Table 1). However, as $\rho \to 0$, all the $\Phi(\rho)$ and $w(\rho)$ curves coincide with the analytic solution of Eq. (13) at $\varepsilon = 0$.

Thus, using the phase functions allows uniquely defining the solutions for the eigenfunctions of stationary bound fermion states in the fields of classical black holes.

## 3. Stationary solutions of second-order equations for scalar particles in the fields of classical black holes

For simplicity in what follows, we consider scalar particles to be massive uncharged particles with zero spin. After separating the variables, the second-order equation for the radial wave functions takes the form of Eq. (1) with the effective potential obtained in [9]:

$$\frac{d^2 \psi^{sc}}{d\rho^2} + 2\left(E_{Schr} - U_{eff}^{sc}\right)\psi^{sc} = 0. \tag{36}$$

For the considered black holes, the function $\omega_{sc}(\rho, \varepsilon)$, which depends on the particle energy),** appears in the effective potentials (see (2) - (5) with $\alpha_{em} = 0$).

If $\omega_{sc}(\rho_\pm, \varepsilon) = 0$, then, for all black holes in a neighborhood of the event horizon, there is a potential well of the form

$$U_{eff}^{sc}\big|_{\rho \to \rho_\pm} = -\frac{1}{8}\frac{1}{(\rho - \rho_\pm)^2}. \tag{37}$$

The coefficient 1/8 is border-line: at this value, particles still do not "fall" on the event horizons. The energies of stationary states can be calculated using formulas (7) - (10) with $\alpha_{em} = 0$.

The asymptotics of the solutions of Eq. (36) as $\rho \to \infty$ are the same at those for Eq. (1) (see (11)).

As $\rho \to \rho_\pm$, the defining equation for solutions (36) with the leading singularity (37) takes the form

$$s^2 - s + \frac{1}{4} = 0, \quad s_{1,2} = \frac{1}{2}. \tag{38}$$

The asymptotics of Eq. (36) then become

$$\psi^{sc}\big|_{\rho \to \rho_\pm} = A_3 \ln|\rho - \rho_\pm| f_2(|\rho - \rho_\pm|) + A_4 f_1(|\rho - \rho_\pm|). \tag{39}$$

---

** It is a parameter of the Schwarzschild metric.

To determine the form of the functions $f_1(|\rho - \rho_\pm|), f_2(|\rho - \rho_\pm|)$, we refer to the model equation.

### 3.1 Second-order model equation

We consider the equation

$$\frac{d^2\psi^{sc}}{d\rho^2} + \frac{1}{4\rho^2}\psi^{sc} + (\varepsilon^2 - 1)\psi^{sc} = 0. \tag{40}$$

The asymptotics in (11) and (39) with the replacement of $|\rho - \rho_\pm|$ by $\rho$ apply to this equation.

In case $\varepsilon = 0$ (the Schwarzschild metric, see (10)), Eq. (40) has the analytical solution

$$\psi^{sc}(\rho) = A_3 \sqrt{\rho} K_0(\rho) + A_4 \sqrt{\rho} I_0(\rho), \tag{41}$$

where $I_0(\rho)$ and $K_0(\rho)$ are the Bessel and Macdonald functions with the asymptotics

$$\sqrt{\rho} K_0(\rho) : \begin{cases} e^{-\rho}, & \rho \to \infty, \\ -\rho^{1/2} \ln \rho, & \rho \to 0, \end{cases} \tag{42}$$

$$\sqrt{\rho} I_0(\rho) : \begin{cases} e^{\rho}, & \rho \to \infty, \\ \rho^{1/2}, & \rho \to 0. \end{cases} \tag{43}$$

We see that if $A_4 = 0$ in Eq. (41), then the solution for the stationary bound state $\varepsilon = 0$ remains that decreases exponentially as $\rho \to \infty$ and is proportional to $-\rho^{1/2} \ln \rho$ as $\rho \to 0$.

In the case $\varepsilon \neq 0$ (see (7) - (9)), the solutions of Eq. (40) have asymptotics (11) as $\rho \to \infty$ and (42), (43) as $\rho \to 0$. Obviously, there are also eigenfunctions with the asymptotics $\sim -\rho^{1/2} \ln \rho$ as $\rho \to 0$ and an exponential decrease as $\rho \to \infty$ for the stationary bound states with energies (7) - (9). With reasonable confidence, we can make a similar conclusion for the solutions of Eq. (36). As in Sec. 2.1, for the final proof of this conclusion and for solving the problem of selecting solutions with physical asymptotics, we use the phase functions (see (17) - (21)) in numerical calculations.

### 3.2 Asymptotics of the functions $\Phi^{sc}(\rho), P^{sc}(\rho)$ and $\psi^{sc}(\rho)$

The asymptotics of the solutions of Eqs. (20), (21) with the potential $U_{eff}^{sc}$ for the scalar particles are given in Table 2. Asymptotics (11), (39) for Eq. (36) are also presented.

The asymptotics in the third and the fourth rows of Table 2 are obtained using expansions (39), (41) of the Bessel and Macdonald functions in a neighborhood of the event horizon and using expressions (17) - (19) for the phase function.

Table 2. Asymptotics of Eqs. (36), (20), (21) with $U_{eff}^{sc}$.

| № | | $\Phi^{sc}(\rho)$ | $(-1)^k \sin\Phi^{sc}(\rho)$ | $P^{sc}(\rho)$ | $\psi^{sc}(\rho)$ |
|---|---|---|---|---|---|
| 1 | $\rho \to \infty$ | $-\arctan\dfrac{1}{\sqrt{1-\varepsilon^2}} + k\pi$ | $-\dfrac{1}{\sqrt{2-\varepsilon^2}}$ | $C_1'\chi_1^{sc}(\rho)e^{-\sqrt{1-\varepsilon^2}\rho}$ | $C_1\chi_1^{sc}(\rho)e^{-\sqrt{1-\varepsilon^2}\rho}$ |
| 2 | $\rho \to \infty$ | $+\arctan\dfrac{1}{\sqrt{1-\varepsilon^2}} + k\pi$ | $+\dfrac{1}{\sqrt{2-\varepsilon^2}}$ | $C_3'\chi_2^{sc}(\rho)e^{\sqrt{1-\varepsilon^2}\rho}$ | $C_2\chi_2^{sc}(\rho)e^{\sqrt{1-\varepsilon^2}\rho}$ |
| 3 | $\rho \to \rho_\pm$ | $\arctan\dfrac{2|\rho-\rho_\pm|(a-\ln|\rho-\rho_\pm|)}{a-2-\ln|\rho-\rho_\pm|} + k\pi$, $a = 0.11593...$ | $\dfrac{2|\rho-\rho_\pm|\times(a-\ln|\rho-\rho_\pm|)}{a-2-\ln|\rho-\rho_\pm|}$ | $A_3'\dfrac{a-2-\ln|\rho-\rho_\pm|}{2\sqrt{|\rho-\rho_\pm|}}$ | $-A_3\sqrt{|\rho-\rho_\pm|}\times\ln|\rho-\rho_\pm|$ |
| 4 | $\rho \to \rho_\pm$ | $\arctan 2|\rho-\rho_\pm| + k\pi$ | $2|\rho-\rho_\pm|$ | $-A_4'\dfrac{1}{2\sqrt{|\rho-\rho_\pm|}}$ | $-A_4\sqrt{|\rho-\rho_\pm|}$ |
| 5 | $\rho \to 0$ | $\dfrac{2}{3}\rho + k\pi$ | $\dfrac{2}{3}\rho$ | $A_5'\rho^{1/2}$ | $A_5\rho^{3/2}$ |
| 6 | $\rho \to 0$ | $-2\rho + k\pi$ | $-2\rho$ | $A_6'\rho^{-3/2}$ | $A_6\rho^{-1/2}$ |

The first, third, and the fifth rows correspond to the asymptotics of solutions of Eqs. (36) for stationary bound states with energies of (7) - (10). The asymptotics in the first and third rows correspond to the domain of the wave functions $\rho \in [\rho_+, \infty)$; the asymptotics in the third and fifth rows correspond to the renge $\rho \in (0, \rho_-]$.

### 3.3 Singular points of the equations for the phase function $\Phi^{sc}(\rho)$

As in the foregoing (see Sec. 2.2.2 for fermions), Eq. (20) with $U_{eff}^{sc}$ has the following singular points on the phase plane $(\rho, \Phi^{sc})$: $(\infty, \Phi_k^\pm)$, $(\rho_\pm, \Phi_k^{(1,2)})$, and $(0, \Phi_k^{(3,4)})$. The characteristics of the singular points for fermions and for scalar particles as $\rho \to \infty$ are the same. The singular points $(\rho_\pm, \Phi_k^{(1,2)})$ in neighborhoods of the event horizons differ in their dependence on $\rho$ as $\rho \to \rho_\pm$. However, the character of the singular points remains the same: the integral curve entering the "node" part of the singular point is

$$\Phi_k^{(1)} = \arctan\frac{2|\rho-\rho_\pm|(a-\ln|\rho-\rho_\pm|)}{a-2-\ln|\rho-\rho_\pm|} + k\pi, \quad k = 0, \pm 1, \pm 2, ... ,$$

and the separatrix separating the "saddle" part of the singular point from the "nod" part is

$$\Phi_k^{(2)} = 2|\rho-\rho_\pm| + k\pi.$$

As in Sec. 2.2.2, the "node" curves $\Phi_k^{(1)}$ are separated from the separatrix $\Phi_k^{(2)}$ at any value of $|\rho-\rho_\pm|$.

*3.4 Numerical solutions of the equations for phase functions of $\Phi^{sc}(\rho)$*

The strategy of integrating Eqs. (20), (21) with the potential $U_{eff}^{sc}$ is the same as in Sec. 2.2.6. The numerical solution completely reproduces the analytic solution of Eq. (40) with $\varepsilon = 0$.

The results of solving Eqs. (20), (21) with the potential $U_{eff}^{sc}$ for the stationary bound states with the energies (7) - (9) qualitatively coincide with the appropriate results for fermions [2] - [4].

The wave functions of scalar particles vanish at the event horizons. As in the fermion case, the scalar particles are located in a neighborhood of the event horizons with an overwhelming probability.

## 4. Stationary solutions of second-order equations for photons in the fields of classical black holes

After separating the variables, the second-order equation for the photon radial wave function becomes [9]

$$\frac{d^2\psi^{ph}}{d\rho^2} + 2\left(E_{Schr}^{ph} - U_{eff}^{ph}\right)\psi^{ph} = 0. \tag{44}$$

For photons, $E_{Schr}^{ph} = \varepsilon^2/2$ because of the absence of the rest mass. The stationary states with energies (7), (8) are possible only for rotating black holes (the Kerr and Kerr-Newman metrics). Because of the absence of the rest mass, the stationary bound states of photons with solutions exponentially decreasing as $\rho \to \infty$ (see Eq. (44)) are impossible. Only photon stationary states of the continuous spectrum with energies of (7), (8) are possible for rotating black holes. The continuous spectrum states for particles with different spins in the fields of the considered black holes will be studied in the forthcoming paper.

## 5. Results

In numerically integrating second-order equations of type (1), (36), we face the problem of selecting the solutions with physical symptotics.

The asymptotics of solutions of Eq. (1) for fermions as $\rho \to \infty$, $\rho \to \rho_{\pm}$, $\rho \to 0$ are given by

$$\psi|_{\rho \to \infty} = C_1 \chi_1(\rho) e^{-\sqrt{1-\varepsilon^2}\rho} + C_2 \chi_2(\rho) e^{\sqrt{1-\varepsilon^2}\rho}, \tag{45}$$

$$\psi|_{\rho \to \rho_{\pm}} = C_3 f_1(|\rho - \rho_{\pm}|) + C_4 f_2(|\rho - \rho_{\pm}|), \tag{46}$$

$$\psi|_{\rho \to 0} = C_5 \varphi_1(\rho) + C_6 \varphi_2(\rho). \tag{47}$$

The asymptotics of solutions of Eq. (36) for scalar particles are

$$\psi^{sc}\big|_{\rho\to\infty} = C_1\chi_1^{sc}(\rho)e^{-\sqrt{1-\varepsilon^2}\rho} + C_2\chi_2^{sc}(\rho)e^{\sqrt{1-\varepsilon^2}\rho}, \qquad (48)$$

$$\psi^{sc}\big|_{\rho\to\rho_\pm} = A_3\ln|\rho-\rho_\pm|f_1^{sc}(|\rho-\rho_\pm|) + A_4 f_2^{sc}(|\rho-\rho_\pm|), \qquad (49)$$

$$\psi^{sc}\big|_{\rho\to 0} = A_5\varphi_1^{sc}(\rho) + A_6\varphi_2^{sc}(\rho). \qquad (50)$$

In (46), (47) and (49), (50), $f_1$, $f_1^{sc}$, $\varphi_1$, and $\varphi_1^{sc}$ are solutions with the asymptotics of the scalar particle and fermion wave functions of stationary bound states, and $f_2$, $f_2^{sc}$, $\varphi_2$, and $\varphi_2^{sc}$ correspond to the solutions with the asymptotics that are not physical in the problem of finding stationary bound states. The coefficients $C_1 \div C_6$ and $A_3 \div A_6$ are integration constants.

The point $\rho=\infty$ is regular, and the points $\rho=\rho_\pm$ and $\rho=0$ are "node" regular singular points, with integral curves entering them.

We consider the domain of the wave functions $D_1$ with $\rho\in[\rho_+,\infty)$. In finding the wave functions of stationary bound states of scalar particles and fermions with energies (7) - (10), taking the "node" character of the singular point $\rho=\rho_+$ into account, we have to integrate "from right to left", choosing the asymptotics with $C_1$ in (45). But an "admixture" of the second asymptotics in (45) with the coefficient $C_2$ appears unavoidably in numerical calculations at large but finite $\rho=N$. For fermions, it is difficult to prove that no "admixture" of a nonphysical solution (for the problem under consideration) appears in (46) with a coefficient $C_4$ in the end of integration as $\rho\to\rho_+$. Similarly, for scalar particles, there are arguments regarding the "admixture" in the numerical calculations of the nonphysical solution in (49) with a coefficient $A_4$.

The situation is even worse for the domain of $D_2$ with $\rho\in(0,\rho_-]$[††]. At both ends of integration, the points $\rho=0$ and $\rho=\rho_-$ are "node" singular points with the integral curves entering them. In this case, the stable process of numerical integration is not provided either by the "right-to-left" or by the "left-to-right" integration.

These difficulties disappear when passing to Eqs. (20), (21) for the phase functions $\Phi(\rho)$. The point $\rho=\infty$ becomes a regular singular point.

For the exponentially decreasing solutions with the coefficient $C_1$ in (45), the singular point $\rho=\infty$ is a "saddle",

---

[††] This region is realized for the Reissner-Nordström metric.

$$\Phi_k^-\big|_{\rho\to\infty} = -\arctan\frac{1}{\sqrt{1-\varepsilon^2}} + k\pi. \tag{51}$$

For the exponentially growing solutions with the coefficient $C_2$ in (45), the singular point $\rho = \infty$ is a "node".

$$\Phi_k^+\big|_{\rho\to\infty} = +\arctan\frac{1}{\sqrt{1-\varepsilon^2}} + k\pi. \tag{52}$$

The singular points $\rho = \rho_\pm$, $\rho = 0$ are irregular, they combine singularities of the "node" and "saddle" types. Naturally, the separatrix separating the "saddle" neighborhood of the singular point from the "node" integral curves does not cross the latter at any $\rho \neq \rho_\pm$, $\rho \neq 0$.

For $D_1$, we start the "right to left" integration from the "saddle" part of the singular point with separatrix (51) and terminate it at the "node" part of the singular point $\rho = \rho_+$ with the asymptotics $f_1(\rho - \rho_+)$ for fermions and $\ln(\rho - \rho_+) f_1^{sc}(\rho - \rho_+)$ for scalar particles (see (46) and (49)). We start the "left-to-right" integration from the "saddle" part to the singular point along the separatrix separating the "saddle" from the "node" at the irregular singular point $\rho = \rho_+$ and terminate it at the "node" with the nonphysical (for the given problem) solutions exponentially growing as $\rho \to \infty$.

For $D_2$, we start the "left-to-right" integration from the "saddle" part of the singular point along the separatrix separating the "saddle" from the "node" at the irregular singular point $\rho = 0$. The integration terminates at the "node" part of the singular point $\rho = \rho_-$ with the asymptotics $f_1(\rho_- - \rho)$ for fermions and $\ln(\rho_- - \rho) f_1^{sc}(\rho_- - \rho)$ for scalar particles.

All three described procedures of numerical integration are stable.

## 6. Conclusions

The Prüfer transformation involving phase functions is a reliable and effective method of numerical integration of the second-order equations for fermions and scalar particles, allowing the solutions with physically acceptable asymptotics to be selected.

### Aknowledgments

The authors thank M.N.Smolyakov for fruitful discussions and discussions. The authors also thank A.L.Novoselova for the essential technical assistance in the paper preparation.